\documentclass[10pt]{article}
\usepackage{graphicx}
\usepackage{amssymb}

\def\beq{\begin{equation}}
\def\eeq{\end{equation}}
\def\ba{\begin{eqnarray}}
\def\ea{\end{eqnarray}}
\def\ar{\begin{array}}
\def\ear{\end{array}}
\def\pa{\partial}

\def\nn{\nonumber\\}

\def\ga{\gamma}

\def\a{\alpha}
\def\b{\beta}

\def\la{\lambda}

\def\vp{\varphi}
\def\ve{\varepsilon}

\def\ri{{\rm i}}
\def\GeV{{\rm GeV}}
\def\eV{{\rm eV}}

\def\a{\alpha}
\def\da{{\dot\alpha}}
\def\b{\beta}
\def\db{{\dot\beta}}
\def\g{\gamma}

\def\ve{\varepsilon}
\def\vp{\varphi}

\def\bn{{\bar\nu}}
\def\bN{{\bar{N}}}

\def\ri{{\rm i}}

\def\cJ{{\cal J}}
\def\cO{{\cal O}}
\def\cC{{\cal C}}
\def\cD{{\cal D}}
\def\cF{{\cal F}}
\def\mL{{\cal {L}}}

\def\sp{\not{\!p}}

\def\sq{\not{\!q}}

\def\sl{\not{l}}

\def\spa{\not{\!\partial}}

\begin{document}
\centerline{\bf\Large Neutrinos, Axions and Conformal Symmetry}

\vspace{5mm}
\centerline{\bf Krzysztof A. Meissner${}^{1}$ and Hermann Nicolai${}^2$}

\vspace{5mm}
\begin{center}
{\it ${}^1$ Institute of Theoretical Physics, University of Warsaw,\\
Ho\.za 69, 00-681 Warsaw, Poland\\ ${}^2$ Max-Planck-Institut f\"ur
Gravitationsphysik (Albert-Einstein-Institut),\\ M\"uhlenberg 1,
D-14476 Potsdam, Germany}
\end{center}

\begin{abstract}

\footnotesize{We demonstrate that radiative breaking of conformal
symmetry (and simultaneously electroweak symmetry) in the Standard
Model with right-chiral neutrinos and a minimally enlarged scalar
sector induces spontaneous breaking of lepton number symmetry, which
naturally gives rise to an axion-like particle with some unusual
features. The couplings of this `axion' to Standard Model particles,
in particular photons and gluons, are entirely determined (and
computable) via the conformal anomaly, and their smallness turns out
to be directly related to the smallness of the masses of light
neutrinos.}
\end{abstract}

\section{Introduction}
It has been known for some time that classically unbroken conformal
symmetry may provide a possible mechanism towards explaining the
stability of the weak scale w.r.t  to the Planck scale
\cite{Bardeen}. In such a scheme all observed mass scales must arise
from a single scale via the quantum mechanical breaking of conformal
invariance induced by the Coleman-Weinberg (CW) effective potential
\cite{CW}. In \cite{MN1} (see \cite{ST,Shap} for an earlier, but
different proposal, and also
\cite{CW1,CW2} for subsequent work along these lines), we have recently
shown that a minimal extension of the Standard Model (SM) with right-chiral
neutrinos and one extra scalar can realize this possibility, provided
that \cite{MN1,MN2}
\begin{itemize}
\item {there are no intermediate mass scales between the weak scale
and the Planck scale $M_P$; and}
\item{the RG evolved couplings exhibit neither Landau poles
nor instabilities over this whole range of energies.}
\end{itemize}
While the first point concerns an issue that must be decided experimentally, 
the second assumption is motivated by the expectation that {\em any}
extension of the SM (with or without supersymmetry) that stays within
the framework of quantum field theory will eventually fail, and that
the main task is therefore to delay the onset of this breakdown to the
Planck scale, where a proper theory of quantum gravity is expected to
replace quantum field theory. This requirement leads to important
restrictions on the SM parameters, which can in principle be tested at LHC.

In this Letter, we study an extension of the model \cite{MN1} with the
usual Higgs doublet $\Phi(x)$, and one extra (weak singlet) scalar field
$\phi(x)$, but with the main difference that this extra scalar field is
now taken to be {\em complex}. Writing
\beq\label{phi}
\phi(x) = \vp (x) \exp \left(\frac{ia(x)}{\sqrt{2}\mu}\right)
\eeq
with real fields $\vp(x)$ and $a(x)$, we will show that $\vp(x)$ can
acquire a non-vanishing vacuum expectation value via radiative corrections.
The field $a(x)$ then gives rise to a (pseudo-)Goldstone particle associated
with the spontaneous breaking of a new {\em global} $U(1)_L$ (modified
lepton number) symmetry. A main new result of the present work is that
this boson, commonly referred to as the `Majoron' \cite{Peccei}, shares
many properties with the axion \cite{axion,WW}. We here explore some of
these features, which are mainly due to `neutrino mediation', especially
the effective couplings of this `axion' to photons and gluons. The latter
may have important implications with regard to the potential suitability
of the axion as a dark matter candidate and for solving the strong CP problem.
Details concerning such phenomenological applications will be given elsewhere.

Apart from its compatibility with the known SM phenomenology, and its
implications for the hierarchy problem, the main virtue of the present
proposal is that it provides a {\em single} source of explanation for
axion couplings and neutrino masses via the conformal anomaly, thereby
tying together in a most economical manner features of the SM previously
thought to be unrelated. If it should turn out that there are indeed
no new scales beyond the weak scale it could thus offer an attractive
and economical alternative to MSSM type models, because low energy
supersymmetry may not be needed to stabilize the weak scale with regard
to the Planck scale if the two conditions stated above are met~\cite{MN2}
(see also \cite{Shap} for a very similar point of view).

\section{Lagrangian}
We refer to \cite{EPP,Pok} for basic properties of the SM, and here only
quote the relevant interaction terms in the (classically conformal)
lagrangian, {\em viz.}
\ba\label{L}
\mL_{\rm int}\!\!\!&=&\!\!\!
\Big(\overline{L}^i\Phi Y_{ij}^E E^j  +
\overline{Q}^i\Phi Y_{ij}^D D^j+\overline{Q}^i\ve\Phi^* Y_{ij}^U U^j\nn
&& + \, \overline{L}^i\ve\Phi^\ast Y_{ij}^\nu N^j+\phi N^{i T}
\cC^{-1}Y^M_{ij} N^j+{\rm h.c.}\Big) \nonumber\\
&& \!\!\!\!\!\!\!\!\!\!\!\!\!\!\!
- \frac{\la_1}{4}\left(\Phi^\dagger\Phi\right)^2
-\frac{\la_2}{2}\left(\Phi^\dagger\Phi\right)\left(\phi^\dagger\phi\right)
-\frac{\la_3}{4}\left(\phi^\dagger\phi\right)^2\,.
\ea
Here $Q^i$ and $L^i$ are the left-chiral quark and lepton doublets,
$U^i$ and $D^i$ the right-chiral up- and down-like quarks, while
$E^i$ are the right-chiral electron-like leptons, and $N^i\equiv\nu^i_R$
the right-chiral neutrinos. As in \cite{MN1} we suppress all indices
except the family indices $i,j=1,2,3$. One can use global
redefinitions of the fermion fields to transform the Yukawa matrices
$Y_{ij}^E$, $Y_{ij}^U$ and $Y_{ij}^M$ to real diagonal matrices.
Furthermore, $Y^D= K_D M_D$ where $K_D$ is a CKM matrix
and $M_D$ is real diagonal. Using the remaining freedom we can then
set $Y^\nu=K_\nu M_\nu U_N$, where $K_\nu$ is a CKM-like matrix
({\em i.e.} with three angles and one phase), $M_\nu$ is real
diagonal and $U_N$ is a unitary matrix with $\det U_N=1$.

Besides the standard (local) $SU(3)_c\times SU(2)_w \times U(1)_Y$
symmetries, the lagrangian (\ref{L}) admits two {\em global} $U(1)$
symmetries. One is the standard baryon number symmetry $U(1)_B$,
\beq
Q^i \to e^{\ri\b} Q^i,\; U^i \to e^{\ri\b} U^i,\; D^i \to e^{\ri\b} D^i;
\eeq
the other is the modified lepton number symmetry $U(1)_L$:
\beq
L^i \to e^{\ri\a} L^i,\  E^i \to e^{\ri\a} E^i,\ N^i \to
e^{\ri\a} N^i,\  \phi \to e^{-2\ri\a}\phi
\eeq
with associated Noether current
\beq\label{JL}
\cJ^\mu_L = \overline{L}^i\ga^\mu L^i
+\overline{E}^i\ga^\mu E^i+\overline{N}^i\ga^\mu N^i
-2 i \phi^\dagger\! \stackrel{\leftrightarrow}{\pa^\mu} \!\phi
\eeq
Notice that the first three terms add up to a purely vectorlike current
of the standard form $= \bar{e}^i\ga^\mu e^i + \bar{\nu}^i\ga^\mu\nu^i$. 
An important feature is that the scalar field $\phi$ also carries lepton 
charge, which can thus `leak' from the fermions via the last term in 
(\ref{JL}); see e.g. \cite{Pelt} and references therein for a discussion 
of this issue.

\section{Minimization of effective potential.}
While the classical potential (\ref{L}) does not induce spontaneous
symmetry breaking, the effective one-loop potential computed from
(\ref{L}) can develop non-vanishing vacuum expectation values for
the scalar fields. This potential is given by the sum of eqns.~(3)
(now with $N\!=\!4$ and $M\!=\!2$) and (6) of \cite{MN1}, to which we
also add the contribution from the $SU(2)_w$ gauge fields which was
not taken into account in \cite{MN1}; explicitly,
\ba\label{EffPot}
V_{{\rm eff}} (H,\varphi) &=&
\frac{\la_1 H^4}{4}+\frac{\la_2
H^2\varphi^2}{2}+\frac{\la_3\varphi^4}{4}
+\frac9{16} \a_w^2 H^4 \ln\left[ \frac{H^2}{v^2}\right] \nn
&&   + \; \frac{3}{256 \pi^2}\big( \la_1 H^2 + \la_2\varphi^2 \big)^2
  \ln \left[ \frac{\la_1 H^2 + \la_2 \varphi^2}{v^2} \right] \nn
&& + \; \frac{1}{256\pi^2}
\big( \la_2 H^2 + \la_3\varphi^2 \big)^2
  \ln \left[ \frac{\la_2 H^2 + \la_3 \varphi^2}{v^2} \right] \nn
&&   + \; \frac1{64\pi^2} F_+^2 \ln \left[\frac{F_+}{v^2}\right]
  + \, \frac1{64\pi^2} F_-^2 \ln \left[\frac{F_-}{v^2} \right]\nn
&&   - \, \frac6{32\pi^2} g_t^4 H^4 \ln\left[ \frac{H^2}{v^2}\right]
- \, \frac1{32\pi^2} Y_M^4 \varphi^4 \ln \left[\frac{\varphi^2}{v^2}\right]
\ea
where $\alpha_w \equiv g_2^2/4\pi$ and $H^2\equiv \Phi^\dagger\Phi$.
$v$ is the mass parameter required by dimensional regularization
(which breaks conformal invariance). Also, we have defined
\ba
F_\pm( H, \varphi) &:=& \frac{3\la_1 + \la_2}4 H^2 +
              \frac{3\la_3 + \la_2}4 \varphi^2  \nn
&& \pm \; \sqrt{ \left[\frac{3\la_1 - \la_2}4 H^2 -
 \frac{3\la_3 - \la_2}4 \varphi^2 \right]^2 + \la_2^2 \varphi^2 H^2}
\ea
As a further simplification, only the contributions from the top quark
(with coupling $g_t$) and one massive neutrino (with coupling $Y_M$)
have been included in (\ref{EffPot}). Next we perform the numerical
minimization subject to the requirements stated in the Introduction.
As in \cite{MN1}, the numerical search shows that there exists a
(small) subset of parameter space compatible with our requirements,
and in particular allowing for the following exemplary set of values:
\beq
\la_1=3.77,\; \la_2=3.72,\; \la_3=3.73,\; g_t=1,\; Y_M^2 = 0.4
\label{exemval}
\eeq
(the approximate $O(6)$ symmetry of the scalar self-couplings is
accidental, and not preserved by quantum corrections, cf.
(\ref{laval}) below). One should note that the natural expansion
parameters in the loop expansion are $\la_i/(4\pi^2)$ so that the
scalar couplings are not very large. For these values, the minimum 
is located at
\beq
\langle H\rangle=2.74\cdot 10^{-5}\, v \; , \quad
\langle\varphi\rangle=1.51\cdot 10^{-4}\,v \;.
\eeq
These values are such that the logarithmic corrections in the
potential are of the same order of magnitude as the tree level
contributions -- as must be the case if the minimum is to be shifted
away from zero. In assessing the reliability of this result one must
therefore ensure that the higher order corrections remain small over
the relevant range of energies in spite of `large logarithms'. Generically,
the latter might invalidate conclusions based on the minimization
of the one-loop potential, but as we will show in a separate publication 
\cite{MN3}, the minimum can nevertheless be stable {\em if} there are 
cancellations among the couplings of the theory ensuring its survival 
up to very large energies -- as is the case for our model (by construction!).
To set the scale for $v$, and thereby for all other dimensionful quantities 
in the model, we impose $\langle H\rangle=174$ GeV. Hence,
\beq\label{H}
\langle H\rangle=174\ \GeV\, ,\;  \langle\varphi\rangle=958\ \GeV \, ,
\; v=3.65\cdot 10^{4}\langle H\rangle
\eeq
After symmetry breaking three degrees of freedom of $\Phi$ are
converted into longitudinal components of $W^\pm$ and $Z^0$, so we
are left with one real scalar field $H$ and the complex field
(\ref{phi}). Defining (at the potential minimum)
\beq
H' = H \cos \beta + \varphi \sin \beta \; , \quad
\varphi' = - H \sin \beta + \varphi \cos \beta
\eeq
the numerical analysis yields the following values (slightly different
from \cite{MN1})
\beq\label{mHiggs}
m_{H'}=207\ \GeV \,,\;\; m_{\varphi'}=477\ \GeV \,,\;\; \sin \beta =0.179
\eeq
while the Goldstone field $a(x)$ stays massless. Note that only the 
components along $H$ of the mass eigenstates couple to the usual 
SM particles. This leads to a clear and testable prediction of the 
present model: the decay amplitudes of $H'$ and $\vp'$ into
SM particles are both proportional to the corresponding decay amplitudes 
of the usual SM Higgs particle with mass values set equal to $m_{H'}$ 
and $m_{\varphi'}$, respectively~\cite{MN1}.

The effective coupling constants are calculated numerically as the respective
fourth-order derivatives of the effective potential at the minimum (now with
the factors $1/(4\pi^2)$ appearing in the loop expansion)
\beq
\frac{\la_1^{{\rm eff}}}{4\pi^2}=0.03665,\;\;
\frac{\la_2^{{\rm eff}}}{4\pi^2}=0.01641, \;\;
\frac{\la_3^{{\rm eff}}}{4\pi^2}=0.02206
\label{laval}
\eeq
Checking the consistency of our basic assumptions now amounts to 
evolving all couplings according 
to their RG equations, with (\ref{laval}) and the known SM couplings 
at the weak scale as the initial values. Performing the steps described 
in \cite{MN1}, and also taking into account the contributions of the 
weak $SU(2)_w$ coupling $\alpha_w$ in the RG evolutions, we have 
verified that for the values indicated above there are indeed no 
Landau poles or instabilities up to the Planck scale.

Although the numbers (\ref{mHiggs}) do not constitute a definitive 
prediction of our model, it turns out that the `window' left open by 
our requirements is fairly small for $m_{H'}$. Preliminary estimates give 
an approximate range $200\,\GeV < m_{H'} < 220\,\GeV$, whereas 
$m_{\vp'}$ can vary over a larger range of values $> \cO(400\, \GeV)$ 
such that the mixing angle $\beta$ decreases with increasing~$m_{\varphi'}$. 
The comparatively large value for $m_{H'}$ distinguishes the present model
from other proposals, and in particular from MSSM type models predicting
$m_{H'} < \cO( 135\,\GeV) $ \cite{Weinberg}.

\section{Neutrino propagators.}
The new effects reported in this Letter all depend crucially on the
mixing of the neutrino degrees of freedom for each neutrino species. 
Inspection of the explicit expressions for the propagators given below 
in (\ref{prop}) shows that this mixing is {\em maximal} in the sense 
that any neutrino degree of freedom can oscillate into any other. 
Because the results would take a much more cumbersome form in terms of 
4-spinors, we temporarily switch to $SL(2,\mathbb{C})$ spinor notation, 
see e.g. \cite{BW} for details, which proves also more convenient for 
the computation of loop diagrams. With $\nu_L \equiv
\frac12(1-\g^5)\nu\equiv \bn^\da$ and $\nu_R \equiv\frac12
(1+\ga^5)\nu \equiv N_\a$, the relevant (free) part of the
Lagrangian reads
\ba
\mL &=& \frac{i}2 \Big( 
  \nu^\a \!\spa_{\a\db} \bn^\db 
      + N^\a \!\!\spa_{\a\db} \bN^\db \Big) + c.c.  \\
&&  \!\!\!\!\!\!\!\!\! + \, m\Big(\nu^\a N_\a + \bn_\da \bN^\da \Big) +
   \frac{M}2 \Big( N^\a N_\a + \bN_\da \bN^\da \Big) \nonumber
\ea
with the Dirac and Majorana mass parameters $m = Y^\nu \langle H \rangle$
and $M= Y^M \langle \vp \rangle$, respectively. For simplicity, we here
consider only one neutrino generation; in the general case the formulas
below will contain additional factors of $Y^\nu Y_M^{-1} Y^\nu$, or traces
over family indices (which may alter our estimates below). Rather than
diagonalize the fields w.r.t to these mass terms, we prefer to work with
{\em non-diagonal propagators}, leaving the fields as they are in the
interaction vertices. The poles of the propagators are obtained via the
standard seesaw formula \cite{PM,seesaw,Yan}
\beq
\mu_\pm^2 = m^2 + \frac12 M^2 \pm \frac12 M^2 \sqrt{1 + \frac{4m^2}{M^2}}
\eeq
whence $\mu_+ \approx M$ and $\mu_- \approx m^2/M \equiv m_\nu$
for $m\ll M$ (so that assuming $|Y_\nu|<10^{-5}$ and substituting
the values (\ref{mHiggs}) found above we get $m_\nu < 1 \,\eV$
\cite{MN1}). Defining $\cD(p):= [(p^2-M^2)(p^2-m_\nu^2)]^{-1}$ we obtain
the propagators (in momentum space)
\ba\label{prop}
\langle \nu_\a\nu_\b \rangle &=& i m^2 M \cD(p)\, \ve_{\a\b}\nn
\langle \nu_\a \bn_\db\rangle &=&
i (p^2 - M^2 - m^2) \cD(p) \sp_{\a\db}
\nn
\langle N_\a N_\b \rangle &=& i Mp^2 \cD(p) \, \ve_{\a\b}
\nn
\langle N_\a  \bN_\db \rangle &=& i (p^2 - m^2) \cD(p) \sp_{\a\db}
\nn
\langle \nu_\a N_\b \rangle &=& i m(p^2-m^2) \cD(p) \ve_{\a\b}  \nn
\langle \nu_\a  \bN_\db \rangle &=& -i mM\cD(p) \sp_{\a\db}
\ea
together with their complex conjugate components (where, as usual,
$\sp_{\a\db} \equiv p_\mu \sigma^\mu_{\a\db}$ and $\ve_{\a\b}$ is the 
$SL(2,\mathbb{C})$ invariant tensor \cite{BW}). It will be essential
for the UV finiteness of the diagrams to be computed below that some 
of these propagators fall off like $\sim p^{-3}$, unlike the standard 
Dirac propagator.


\section{Neutrino triangles and `axion'  vertices.}
With the above propagators we can now proceed to compute various couplings
involving the `axion' $a$, which are mediated by neutrino mixing via two
or three-loop diagrams. In order to extract the new vertices from (\ref{L})
(and to have canonically normalized kinetic terms for these scalar fields),
we set $\mu=\langle\vp\rangle$ in (\ref{phi}) and expand
\beq
\phi(x) = \langle\vp\rangle +
\frac1{\sqrt{2}} \big(\vp(x) + i a(x)\big) + \dots
\eeq
We here only present results for the axion-photon coupling
$aF\tilde{F}$, whose lowest order contribution is given by the
two-loop diagram depicted in Fig.1. Because parity is (in fact,
maximally) broken, there are also `dilatonic' couplings of type
$aFF$, whose determination will require in addition the
consideration of diagrams with photons emanating from the internal
$W$-line. To simplify the calculation we first compute the `blob' in
Fig.~1, which is proportional to the integral (after a Wick rotation)
\ba &&
\int \frac{d^4l}{(2\pi)^4}
\frac{\sl_{\a\db}}{(l^2+ M^2) [(l+q)^2 +M^2]l^2 [(l+p_1)^2 + m_W^2]} \nn
&& \qquad\qquad + \; (p_1 \leftrightarrow p_2)
\ea
where $m_W$ is the $W$-boson mass, and where we set $m\!=\!m_\nu\!
=\!0$ in order to simplify the integrand (a valid approximation because
$m,m_\nu\!\ll \! m_W,M$). Here we are interested in the result for
small axion momentum $q^\mu = p_1^\mu + p_2^\mu$. Putting in all
factors and reverting back to 4-spinor notation, the `form factor'
for the electron-axion vertex for large $p_1,p_2$ is well approximated
by (modulo terms of order $\cO(M^2/p_1^2)$ and $\cO(m_W^2/p_1^2)$)
\beq\label{FF}
\cF (q,p_1) = \frac{m_\nu Y_M\a_w}{8\pi}
\frac{1+\ga^5}2\frac{\sq}{p_1^2 + m_W^2}
\eeq
where now $\sq\equiv \ga^\mu q_\mu$. This
formula makes obvious one of our main assertions, namely that the
effective axionic couplings are proportional to the light neutrino masses,
and vanish in the limit $m_\nu\rightarrow 0$. The complete expression for
the effective axion-electron vertex at small $q^\mu$ will be given
in \cite{MN3}. {\it Inter alia} the above vertex would also
determine the rate of energy loss through radiation of `axions' from
charged plasma in the Sun's core.

In order to arrive at an estimate of the axion-photon coupling, we next
substitute the result (\ref{FF}) into the expression for the electron triangle
shown in Fig.~1. Though superficially similar to the diagram giving rise
to the well known $\ga^5$-anomaly in QED, the integral here is
{\em convergent} because of the damping of the integrand due to
the `form factor' (\ref{FF}). After some computation we obtain
\beq\label{fa}
\mL_{\rm eff}^{a\ga\ga} =
\frac1{4f_a} a \, F^{\mu\nu} \tilde{F}_{\mu\nu} \; ,
\quad f_a = \frac{2 \pi^2 m_W^2}{\a_w \a_{em} m_\nu Y_M}
\eeq
Substituting numbers we find $f_a= \cO( 10^{15} \, \GeV)$ which is
outside the range of existing or planned experiments \cite{OSQAR}.
However, apart from the simplifications introduced above which may
still affect this estimate, one must keep in mind that the
simultaneous presence of $aFF$ couplings may substantially alter the
analysis with regard to observable effects.

The gluonic couplings can be analyzed in a similar way \cite{MN3}.
For their determination we must evaluate the three-loop diagram
shown in Fig.~2, as well as analogous diagrams not shown, with $Z$
boson exchange and a triangle consisting only of neutrino lines, and
with gluons emanating from the quark line connecting two $W$ or $Z$
bosons. This calculation is complicated by the fact that the `quark
box' is dominated by small momenta, where $\a_s$ is large. A very
crude estimate can be derived by replacing $\a_{em}$ in (\ref{fa})
by $6\a_w \a_s/\pi$, which yields
\beq\label{ga}
\mL_{\rm eff}^{agg} =
\frac{\a_s}{8 \pi g_a} a\, G^{\mu\nu} \tilde{G}_{\mu\nu} \; ,
\quad g_a \cong \frac{\pi^2 m_W^2}{6 Y_M \a_w^2 m_\nu}
\eeq
This gives $g_a = \cO(10^{16}\, \GeV)$. Summing over quark flavors,
as well as taking into account all relevant diagrams could bring
this number down, and into the range that could make the axion a
viable dark matter candidate according to standard reasoning (see
e.g. \cite{Mu}, p.~280ff, but the actual numbers are subject to large
uncertainties \cite{Sikivie}). However, before starting the
actual comparison, conventional lines of reasoning must be
re-examined in view of the fact that parity violation now also
allows for `dilatonic' couplings $\propto aGG$, which may vitiate
some of the accepted arguments (for instance, concerning periodicity
properties of the `axion' potential and the location of the minimum
in this potential). Similar caveats may apply to the use of our
`axion' for solving the strong CP problem: although for this purpose
the precise value of $g_a$ does not matter so much, it is not clear
whether the present scenario would lead to $\langle a
\rangle = 0$, as generally required for the solution of the strong
CP problem.

\section{Conclusions and Outlook}
In \cite{MN1} it was shown that all mass scales of the SM can be
generated via radiative corrections and the conformal anomaly
{\em from a single scale} $v$, which is set here by the choice of
$\langle H \rangle$ in eq.~(\ref{H}). In particular, no large scales
beyond the SM are needed to explain the smallness of neutrino masses,
if one allows for the entries of the Yukawa coupling matrix $Y^\nu$
to assume values in the range $\cO(1)$ -- $\cO(10^{-5})$ (this is the
main difference with more conventional seesaw-type scenarios, for which
$m \approx \cO(100 \, \GeV)$ and $M$ must be assumed very large, resulting
in very large masses for the heavy neutrinos, see e.g.~\cite{seesaw1}).
In this Letter, we have extended these considerations to another sector
of the SM, by showing that the replacement of the real field $\vp$ by
the complex scalar $\phi$ makes it possible to also accommodate an
axion-like degree of freedom in the model. Again, the relevant (large)
mass scales emerge rather naturally by radiative corrections, and without
the need to introduce any new large mass scales by hand.

The main novelty of the present work is the proposal to identify the
`Majoron' of \cite{Peccei} with the axion, and to do so in conjunction 
with the quantum mechanical breaking of (classical) conformal invariance. 
This entails several unusual features, for instance the fact that our 
`axion' cannot be assigned a definite parity, unlike the standard 
axion of \cite{WW}. The crucial ingredient here is the maximal neutrino 
mixing in (\ref{prop}), which mediates the axionic couplings (\ref{fa}) 
and (\ref{ga}). Without such neutrino mediation the latter couplings 
would simply be absent if the lepton number symmetry $U(1)_L$ is 
non-anomalous \cite{CNP}. This part of our proposal does not depend 
on the symmetry breaking mechanism, and might thus also work without 
the assumption of classical conformal invariance. The cosmological 
consequences of this scenario (e.g. for leptogenesis in the early 
universe) remain to be explored.

\vspace{0.5cm}
\noindent{\bf {Acknowledgments:}} K.A.M. was partially
supported by the EU grant MRTN-CT-2006-035863 and the Polish grant
N202 081 32/1844. K.A.M. thanks the AEI for hospitality and support
during this work.

\begin{center}
\includegraphics[width=13cm,viewport= 100 610 520 760,clip]{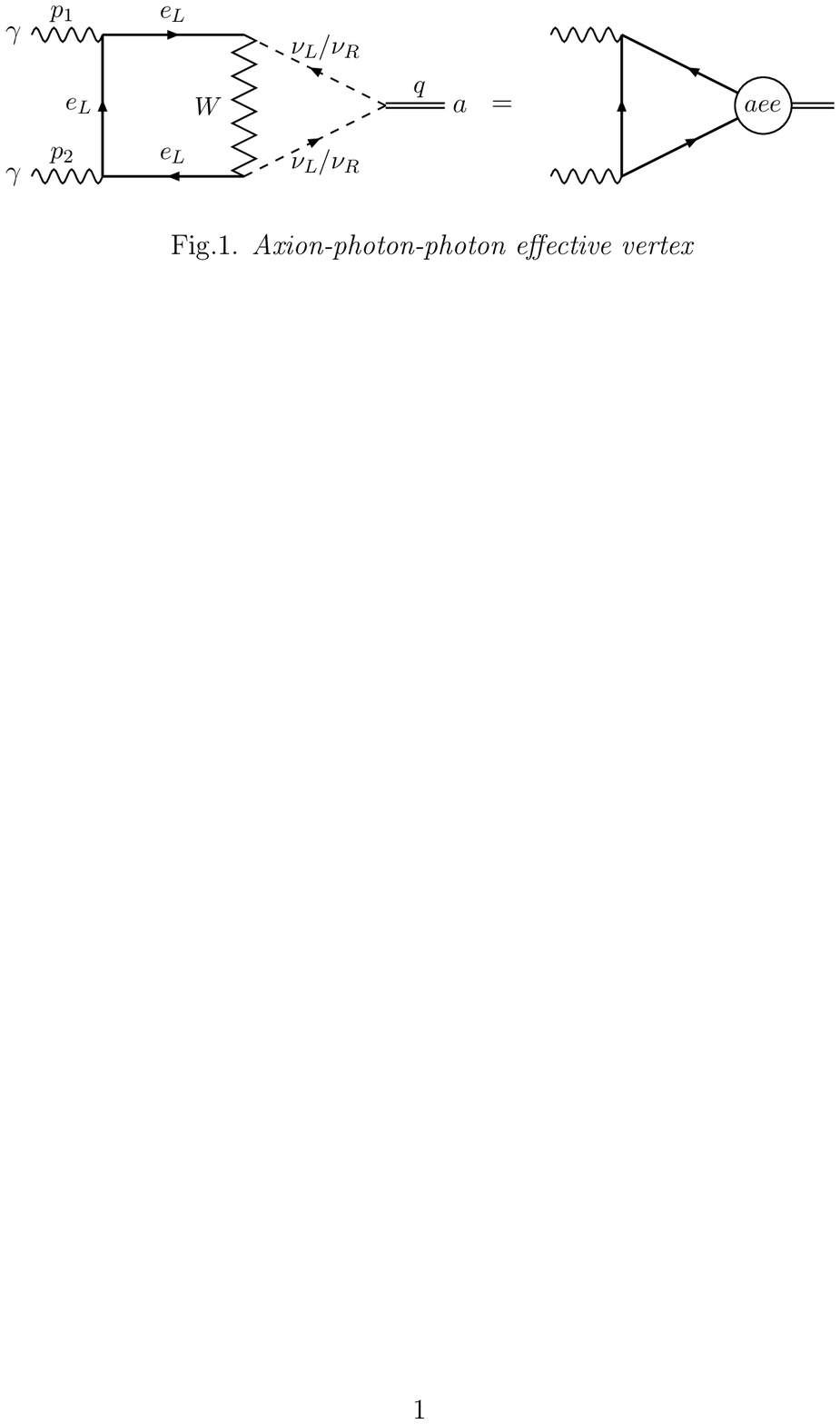}
\includegraphics[width=13cm,viewport= 70 610 490 760,clip]{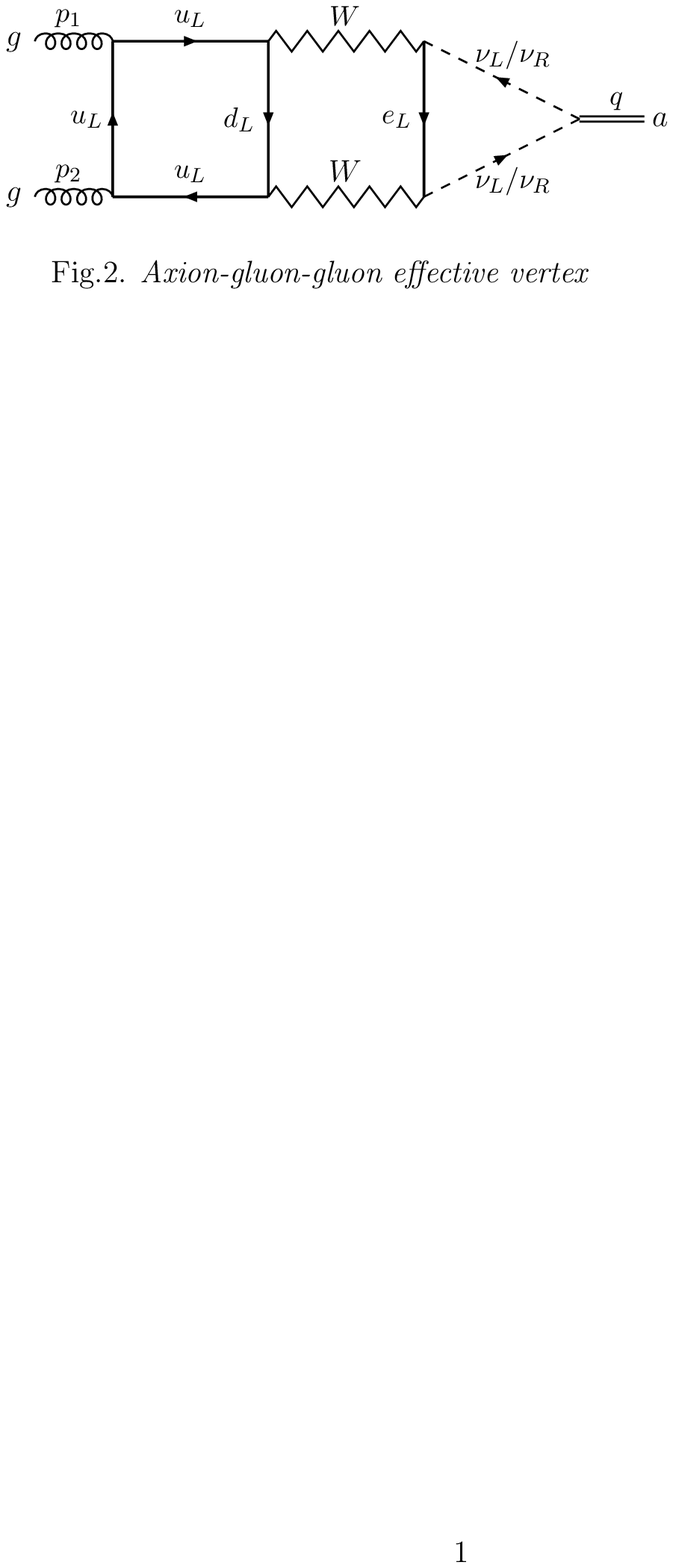}
\end{center}


\begin{thebibliography}{99}

\bibitem{Bardeen} W.A.~Bardeen, FERMILAB-CONF-95-391-T

\bibitem{CW} S.~Coleman and E.~Weinberg,
  Phys. Rev. {\bf D7} (1973) 1888.

\bibitem{MN1} K.A.~Meissner and H.~Nicolai,
  Phys. Lett. {\bf B648} (2007) 312, {\tt hep-th/0612165}.

\bibitem{ST} M.~Shaposhnikov and I.~Tkachev,
  Phys. Lett. {\bf B639} (2006) 414, {\tt hep-ph/0604236}.

\bibitem{Shap} M.~Shaposhnikov,
  arXiv:0708.3550 [hep-th].

\bibitem{CW1} R.~Foot,~A.~Kobakhidze and R.~Volkas,
  Phys.Lett. {\bf B655} (2007) 156, {\tt arXiv:0704.1165[hep-ph]}.

\bibitem{CW2} R.~Foot,~A.~Kobakhidze,~K.L.McDonald,~R.~Volkas,
  Phys.Rev. {\bf D77}:035006 (2008), {\tt arXiv:0709.2750}.

\bibitem{MN2} K.A.~Meissner and H.~Nicolai,
  Phys.Lett. {\bf B660} (2008) 260, {\tt arXiv:0710.2840[hep-th]}.

\bibitem{Peccei} Y.~Chikashige, R.N.~Mohapatra and R.D.~Peccei,
   Phys. Rev. Lett. {\bf 45} (1980) 1926

\bibitem{axion} R.D.~Peccei and H.R.~Quinn, Phys. Rev. Lett.
      {\bf 38} (1977) 1440;

\bibitem{WW} S.~Weinberg, Phys. Rev. Lett. {\bf 40} (1978) 223; F.~Wilczek,
   Phys. Rev. Lett. {\bf 40} (1978) 279.


\bibitem{EPP}
 O.~Nachtmann, {\it Elementary Particle Physics: Concepts
   and Phenomena}, Springer Verlag (1999).

\bibitem{Pok} S.~Pokorski, {\it Gauge Field Theories}, Cambridge Univ. Press,
   2nd edition (2000).

\bibitem{Pelt} J.T.~Peltoniemi, {\tt hep-ph/9511416}

\bibitem{MN3} K.A.~Meissner and H.~Nicolai, {\it Renormalization group and 
   effective potential in classically conformal theories}, to appear.

\bibitem{Weinberg} S.~Weinberg, {\it The Quantum Theory of Fields III:
   Supersymmetry}, Cambridge Univ. Press (2000)

\bibitem{BW} J.~Bagger and J.~Wess, {\it Supersymmetry and Supergravity},
   Princeton University Press, 1984.

\bibitem{PM} P.~Minkowski, Phys. Lett. {\bf B67} (1977) 421

\bibitem{seesaw}
 M.~Gell-Mann, P.~Ramond and R.~Slansky, in
{\it Supergravity}, P. van Nieuwenhuizen and D.Z. Freedman (eds.)
(North-Holland) (1979) 315.

\bibitem{Yan} T.~Yanagida, Prog.Theor.Phys. {\bf 64} (1980) 1103.

\bibitem{OSQAR} P.~Pugnat et al., OSQAR Collaboration, {\tt 
arXiv:0712.3362 [hep-ex]}

\bibitem{Mu} V. Mukhanov, {\it Physical Foundations of Cosmology},
   Cambridge University Press (2005).

\bibitem{Sikivie}
  P. Sikivie,  Lect.\ Notes Phys.\  {\bf 741} (2008) 19
  [arXiv:astro-ph/0610440].

\bibitem{seesaw1} M.~Lindner and W. Rodejohann, JHEP {\bf 0705:}089 (2007),
  {\tt hep-ph/0703171}.

\bibitem{CNP} R.~Chanda, J.F.~Nieves and P.B.~Pal,
   Phys. Rev. {\bf D37} (1988) 2714

\end{thebibliography}
\end{document}